\documentclass[twocolumn,english,aps,prl,twocolumn,superscriptaddress,showpacs,floatfix]{revtex4-1}
\usepackage{mathptmx}

\usepackage[T1]{fontenc}
\usepackage[latin9]{inputenc}
\usepackage{array}
\usepackage{multirow}
\usepackage{amsmath}
\usepackage{amssymb}
\usepackage{graphicx}

\makeatletter

\newcommand{\lyxmathsym}[1]{\ifmmode\begingroup\def\b@ld{bold}
  \text{\ifx\math@version\b@ld\bfseries\fi#1}\endgroup\else#1\fi}

\providecommand{\tabularnewline}{\\}

\makeatother

\usepackage{babel}
\begin{document}
\title{Spin Dynamics of the Centrosymmetric Skyrmion Material GdRu$_{2}$Si$_{2}$}
\author{Joseph A. M. Paddison}
\email{paddisonja@ornl.gov}

\affiliation{Materials Science and Technology Division, Oak Ridge National Laboratory,
Oak Ridge, Tennessee 37831, USA}
\affiliation{Neutron Scattering Division, Oak Ridge National Laboratory, Oak Ridge,
Tennessee 37831, USA}
\author{Juba Bouaziz}
\affiliation{Peter Grünberg Institut and Institute for Advanced Simulation, Forschungszentrum
Jülich and JARA, 52425 Jülich, Germany}
\author{Andrew F. May}
\affiliation{Materials Science and Technology Division, Oak Ridge National Laboratory,
Oak Ridge, Tennessee 37831, USA}
\author{Qiang Zhang}
\affiliation{Neutron Scattering Division, Oak Ridge National Laboratory, Oak Ridge,
Tennessee 37831, USA}
\author{Stuart Calder}
\affiliation{Neutron Scattering Division, Oak Ridge National Laboratory, Oak Ridge,
Tennessee 37831, USA}
\author{Douglas Abernathy}
\affiliation{Neutron Scattering Division, Oak Ridge National Laboratory, Oak Ridge,
Tennessee 37831, USA}
\author{Julie B. Staunton}
\affiliation{Department of Physics, University of Warwick, Coventry CV4 7AL, United
Kingdom}
\author{Stefan Bl\"{u}gel}
\affiliation{Peter Grünberg Institut and Institute for Advanced Simulation, Forschungszentrum
Jülich and JARA, 52425 Jülich, Germany}
\author{Andrew D. Christianson}
\email{christiansad@ornl.gov}

\affiliation{Materials Science and Technology Division, Oak Ridge National Laboratory,
Oak Ridge, Tennessee 37831, USA}
\begin{abstract}
Magnetic skyrmion crystals are traditionally associated with non-centrosymmetric
crystal structures; however, it has been demonstrated that skyrmion
crystals can be stabilized by competing interactions in centrosymmetric
crystals. To understand and optimize the physical responses associated
with topologically-nontrivial skyrmion textures, it is important to
quantify their magnetic interactions by comparing theoretical predictions
with spectroscopic data. Here, we present neutron diffraction and
spectroscopy data on the centrosymmetric skyrmion material GdRu$_{2}$Si$_{2}$,
and show that the key spectroscopic features can be explained by the
magnetic interactions calculated using density-functional theory calculations.
We further show that the recently-proposed 2-$\mathbf{q}$ \textquotedblleft topological
spin stripe\textquotedblright{} structure yields better agreement
with our data than a 1-$\mathbf{q}$ helical structure, and identify
how the magnetic structure evolves with temperature.
\end{abstract}

\maketitle
Magnetic skyrmions are nanometer-scale swirling spin textures that
have attracted renewed interest because of their potential applications
in high-density magnetic memory technologies \citep{Bogdanov_2020,Tokura_2021}.
This possibility is based on two properties of skyrmions. First, their
nontrivial topology offers them a high degree of protection against
external perturbations. Second, magnetic skyrmions can be stabilized
by competing (frustrated) magnetic interactions, which can yield skyrmions
of smaller dimensions compared with those that are stabilized by the
traditional mechanism of antisymmetric exchange interactions \citep{Okubo_2012,Leonov_2015}.
The correspondingly higher skyrmion density can yield larger macroscopic
responses, such as topological Hall signals in itinerant-electron
materials \citep{Kurumaji_2019,Yao_2020}. 

\begin{figure}[h]
\includegraphics{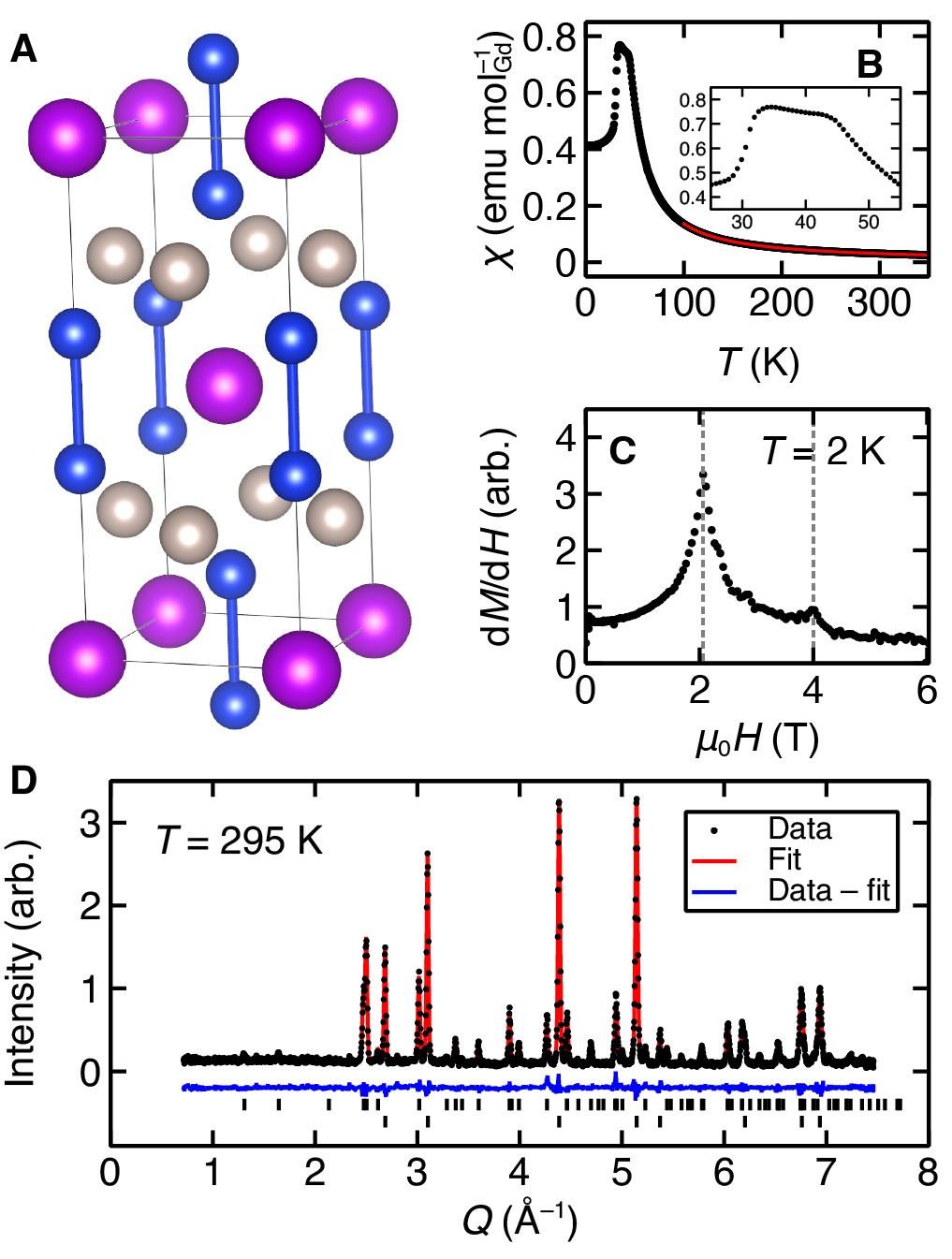}
\centering{}\caption{\label{fig:fig1}\textbf{Structural and magnetic characterization
of }GdRu$_{2}$Si$_{2}$\textbf{.} (a) Crystal structure of GdRu$_{2}$Si$_{2}$,
showing Gd (purple), Ru (grey), and Si (blue) atoms. (b) Magnetic
susceptibility measured in a $1$\ T applied field (black points)
and Curie-Weiss fit (red line), indicating $\mu_{\mathrm{eff}}=7.978(1)\,\mu_{\mathrm{B}}$
per Gd and Curie-Weiss temperature $41.63(2)$\,K. (c) Field derivative
of magnetization at $T=2$\,K showing magnetic phase transitions
in small applied field. (d) Room-temperature neutron-diffraction data
($\lambda=1.536$\,\AA) showing data (black
points), Rietveld fit (red line) and data--fit (blue line). The upper
tick marks identify nuclear peaks from GdRu$_{2}$Si$_{2}$, and the
lower tick marks identify peaks from the Al sample holder. Refined
structural parameters: $a=4.165(1)$\,\AA, $c=9.613(3)$\,\AA,
$z(\textrm{Si})=0.367(5)$, $B(\mathrm{Gd})=0.47(7)$\,\AA$^{2}$,
$B(\mathrm{Ru})=0.39(6)$\,\AA$^{2}$, $B(\mathrm{Si})=0.44(8)$\,\AA$^{2}$.}
\end{figure}

The possibility of stabilizing skyrmions by competing interactions
has expanded the space of materials candidates to include centrosymmetric
systems. Intermetallic materials containing spin-only Gd$^{3+}$ or
Eu$^{2+}$ ions have proved fertile ground for realizing magnetic
skyrmion crystals. Skyrmion phases have been discovered where the
magnetic ions form different lattice geometries, including triangular
in Gd$_{2}$PdSi$_{3}$ \citep{Mallik_1998,Zhang_2020,Sampathkumaran_2000,Kurumaji_2019,Hirschberger_2020a},
breathing kagome in Gd$_{3}$Ru$_{4}$Al$_{12}$ \citep{Hirschberger_2019},
and square in GdRu$_{2}$Si$_{2}$ \citep{Garnier_1995,Garnier_1996,Khanh_2020,Khanh_2022,Rotter_2007,Samanta_2008,Yasui_2020,Eremeev_2023}.
These materials share notably similar magnetic properties. The net
magnetic interaction strength, as measured by the Weiss temperature,
is ferromagnetic. Long-range magnetic ordering is incommensurate with
a periodicity of $\lesssim10$ crystallographic unit cells. Applying
a magnetic field in the magnetically-ordered phase generates a phase
transition to a skyrmion phase---a multi-\textbf{q }structure formed
by superposing sinusoidal or helical modulations with multiple wavevectors
\textbf{q}. 

The similar magnetic properties of centrosymmetric skyrmion materials
suggest that similar interactions are at play; however, the nature
of these interactions remains a matter of debate \citep{Bouaziz_2022,Nomoto_2020,Hayami_2021b,Nomoto_2023}.
While local magnetic moments reside on the Gd$^{3+}$ ions, the magnetic
interactions in these intermetallic materials are primarily mediated
by conduction electrons \citep{Matsuyama_2023}. Consequently, magnetic
interactions are long-ranged, presenting challenges for first-principles
modeling and experimental parametrization alike. Moreover, the energy
resolution of resonant inelastic X-ray measurements is too coarse
to resolve the low-energy ($\sim5$ meV) magnetic signals in these
systems \citep{Paddison_2022}. Therefore, high-quality neutron spectroscopy
measurements are crucial to understand the physical responses associated
with topologically-nontrivial skyrmion textures.

In this article, we present neutron diffraction and spectroscopy data
on the centrosymmetric skyrmion material GdRu$_{2}$Si$_{2}$, and
compare its experimental magnetic excitation spectrum with the predictions
of state-of-the-art theoretical models \citep{Bouaziz_2022}. The
crystal structure of GdRu$_{2}$Si$_{2}$ is tetragonal (space group
$I4/mmm$; $a\approx4.16$\,\AA, $c=9.60$\,\AA) and the
Gd$^{3+}$ ions form a square lattice in the $ab$ plane {[}Figure~\ref{fig:fig1}(a){]}.
We choose GdRu$_{2}$Si$_{2}$ to compare with theory because of the
apparent simplicity of its crystal structure. This contrasts with
Gd$_{2}$PdSi$_{3}$, in which the Pd and Si counter-ions form a superstructure
that significantly complicates its magnetic properties \citep{Tang_2011,Paddison_2022}.
Despite its chemical simplicity, the magnetic behavior of GdRu$_{2}$Si$_{2}$
is subtle. It undergoes two magnetic phase transitions upon cooling
in zero field, at $T_{N}\approx45$\,K and $T^{\prime}\approx38$\,K
\citep{Garnier_1995,Garnier_1996,Yasui_2020}. We will refer to the
$T<T^{\prime}$ state as Phase 1 and the precursor state at $T^{\prime}<T<T_{N}$
as Phase 4, following Ref.~\citep{Wood_2023}. The zero-field structure
below $T_{N2}$ is a 2-\textbf{q }spin texture in which the two wavevectors
have slightly different magnitudes, which is referred to as a ``topological
spin stripe'' state because the topological charge of the spin texture
exhibits a one-dimensional sinusoidal oscillation \citep{Wood_2023}
and resembles theoretically-proposed structures \citep{Ozawa_2016,Hayami_2021}.
In contrast, the magnetic structure of the precursor state is not
yet understood; moreover, the interpretation of the Hall signal is
challenging in this regime due to thermal spin fluctuations at elevated
temperatures \citep{Ishizuka_2018}.

Our study reveals three key results. First, the precursor state at
$T^{\prime}<T<T_{N}$ involves a single magnetic propagation vector
which splits into two wavevectors with different magnitudes at $T^{\prime}$,
hinting that multi-spin magnetic interactions become more significant
on cooling the sample \citep{Mendive-Tapia_2017}. Second, at low
temperature, the key features of our neutron-spectroscopy data are
reproduced with high fidelity by the magnetic interactions obtained
from density-functional theory \citep{Bouaziz_2022}. This result
reveals that the magnetic Hamiltonian is dominated by isotropic interactions.
Third, the agreement with spectroscopic data is improved by assuming
the recently-proposed 2-$\mathbf{q}$ topological spin stripe ground
state \citep{Wood_2023}, compared to a 1-$\mathbf{q}$\textbf{ }helical
ground state \citep{Yasui_2020}, providing strong support for the
former model. Our results show that first-principles calculations
account remarkably well for the magnetic interactions, and place strong
constraints on the magnitude and type of interactions that should
be included in future models.

\begin{figure*}
\begin{centering}
\includegraphics{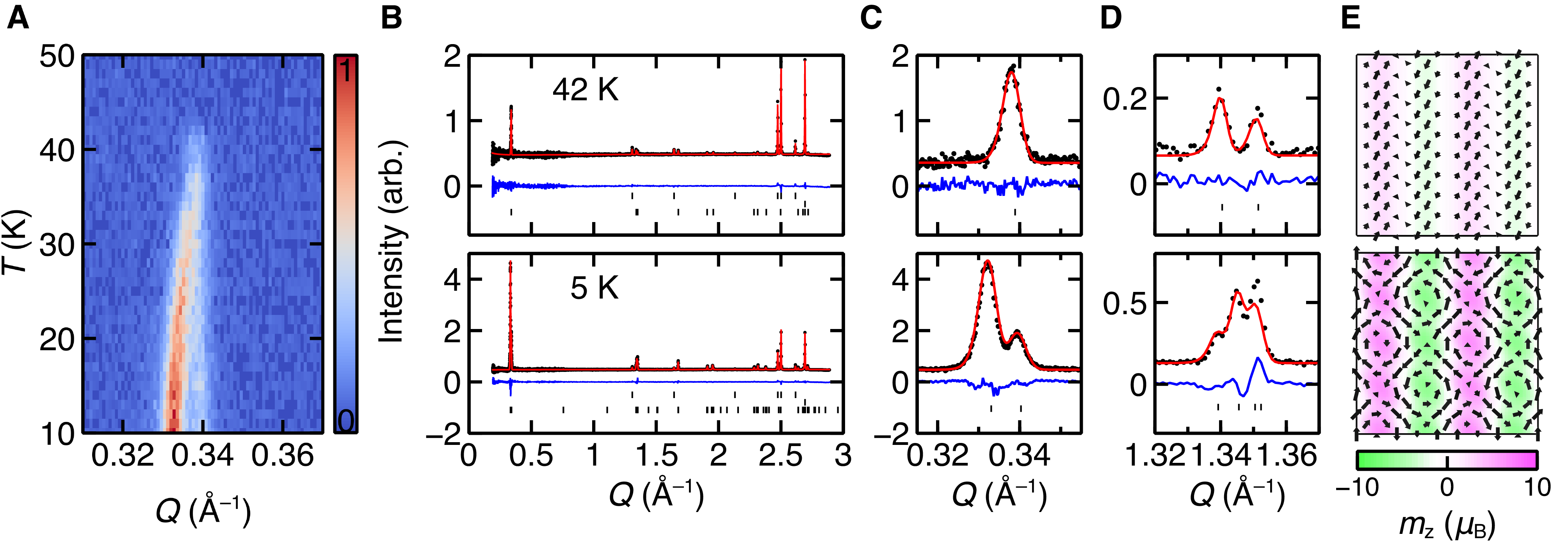}
\par\end{centering}
\centering{}\caption{\label{fig:fig2}\textbf{Neutron diffraction data and Rietveld refinements.}
(a) Dependence of the intensity of the magnetic $(q00)$ reflection,
shown in false color, on wavevector magnitude $Q$ and temperature
$T$. A single magnetic peak appears at $T_{N}\approx45$\,K and
splits into two peaks below $T^{\prime}\approx38$\,K. (b) Neutron
powder diffraction data (black circles), Rietveld fits (red lines),
and data--fit (blue lines) at $T=42$\,K (upper panel) and $5$\,K
(lower panel). Tick marks indicate (top to bottom) nuclear, Al, and
magnetic peaks. (c) Magnetic $(q,0,0)$ peak at $42$\,K (upper panel)
and $(q_{1},0,0)$ and $(q_{2},0,0)$ peaks at $5$\,K (lower panel).
(d) Magnetic $(1-q,0,1)$ and $(q,0,2)$ peaks at $42$\,K (upper
panel), and $(0,1-q_{2},1)$, $(1-q_{1},0,1)$, $(q_{1},0,2)$, and
$(0,q_{2},2)$ peaks at $5$\,K (lower panel). (e) Graphical representation
of the refined 1-$\mathbf{q}$ sinusoidal magnetic structure at $42$\,K
(upper panel) and 2-$\mathbf{q}$ topological spin-stripe structure
including $\mathbf{q}_{1}$, (lower panel). Spin components in the
$ab$ plane are shown as black arrows and the $c$-axis spin component
is shown in false color.}
\end{figure*}

We prepared a polycrystalline sample of $^{160}$GdRu$_{2}$Si$_{2}$
(mass $\sim$$1.3$\,g) by arc melting. The use of isotopically-enriched
$^{160}$Gd (98.1\% enrichment) is essential for successful neutron-scattering
experiments, since isotopically natural Gd contains the strongly neutron-absorbing
isotopes $^{155}$Gd and $^{157}$Gd. Neutron-scattering experiments
were performed at ORNL using a suite of instruments. Room-temperature
neutron diffraction data were measured using the HB-2A diffractometer
($\lambda=1.536$\,\AA). Low-temperature magnetic
diffraction data were measured using the POWGEN diffractometer with
a high measured resolution (full width at half maximum) of $\delta Q/Q=0.015$
at $Q=0.34$\,\AA$^{-1}$. Inelastic measurements were performed
using ARCS spectrometer with incident energies $E_{i}=4$, $8$, and
$14$\,meV, which covers the bandwidth of magnetic excitations. To
reduce the neutron absorption from remaining $^{155}$Gd and $^{157}$Gd,
the sample was loaded in an annular geometry into an Al container
(HB-2A and ARCS) or V container (POWGEN). Cooling was provided by
closed-cycle refrigerators. Our inelastic data were corrected for
the energy dependent neutron absorption, which was significant only
for $E_{i}=4$\,meV, and were placed in absolute intensity units
(bn\,sr$^{-1}$\,meV$^{-1}$ per Gd) by normalization to the nuclear
Bragg profile.

The bulk magnetic susceptibility of our sample is shown in Figure~\ref{fig:fig1}(b).
Our data indicate a magnetic ordering temperature $T_{N}\approx45$\,K,
and a Curie-Weiss temperature of $41.63(2)$\,K, consistent with
previous results \citep{Garnier_1995,Garnier_1996} and with net ferromagnetic
interactions. The bulk susceptibility data show a plateau-like feature
between $\sim$33 and 45\,K {[}inset to Figure~\ref{fig:fig1}(b){]},
which is consistent with the two closely-spaced magnetic phase transitions
at $T_{N}\approx45$\,K and $T^{\prime}\approx38$\,K reported previously
\citep{Garnier_1995,Garnier_1996,Yasui_2020}. Below, we will discuss
the temperature evolution of the zero-field magnetic structure as
this feature is crossed. The derivative of the magnetization with
respect to applied magnetic field is shown in Figure~\ref{fig:fig1}(c),
and shows two field-induced magnetic phase transitions at applied
fields of approximately 2 and 4\,T. Powder neutron-diffraction data
collected at room temperature are shown in Figure~\ref{fig:fig1}(d),
and are quantitatively modeled by the crystal structure shown in Figure~\ref{fig:fig1}(a),
as demonstrated by the high-quality Rietveld fit. 

Having established that the magnetic and crystallographic properties
of our sample are consistent with the literature, we turn to the temperature
evolution of the zero-field magnetic structure. Figure~\ref{fig:fig2}(a)
shows the temperature evolution of the neutron diffraction data measured
using POWGEN. Below $T_{N}$, a new magnetic peak appears at $Q\approx0.34$\,\AA$^{-1}$.
As the sample temperature is reduced below $\approx35$\,K, this
peak appears to broaden, and develops into a second peak at smaller
$Q$. On cooling the sample further, the second peak increases in
intensity and in separation from the initial peak. Profile fitting
of the diffraction data shown in Figure~\ref{fig:fig2}(b--d) reveals
that the data collected in Phase 4 ($T=42$\,K) can be fitted with
a single magnetic propagation vector, $\mathbf{q}=[0.2242(6),0,0]$,
whereas data collected in Phase 1 ($T=5$\,K) require two propagation
vectors of different magnitudes, $\mathbf{q}_{1}=[0.2203(2),0,0]$
and $\mathbf{q}_{2}=[0,0.2251(4),0]$. Since \textbf{$\mathbf{q}\approx\mathbf{q}_{2}$},
these results suggest that the phase transition at $T^{\prime}$ is
associated with the emergence of the additional magnetic propagation
vector $\mathbf{q}_{1}$. 

\begin{table*}
\centering{}%
\begin{tabular}{c|c|cc|cccc|c}
\hline 
$T$ (K) & Structure type & \multicolumn{2}{c|}{$\mathbf{q}$ (r.l.u.)} & $\mu_{\parallel\mathbf{a}}$ ($\mu_{\mathrm{B}}$) & $\mu_{\parallel\mathbf{b}}$ ($\mu_{\mathrm{B}}$) & $\mu_{\parallel\mathbf{c}}$ ($\mu_{\mathrm{B}}$) & $\mathrm{max}(\mu_{\mathrm{ord}})$ ($\mu_{\mathrm{B}}$) & $R_{\mathrm{wp}}$ (\%)\tabularnewline
\hline 
\hline 
$42$ & Sine & \multicolumn{2}{c|}{$[0.2242(6),0,0]$} & $1.00(23)$ & $2.17(13)$ & $2.39(12)$ & $3.38(14)$ & $18.0$\tabularnewline
$42$ & Sine & \multicolumn{2}{c|}{} & $0^{\ast}$ & $2.32(3)^{\dagger}$ & $2.32(3)^{\dagger}$ & $3.28(5)$ & $18.1$\tabularnewline
$42$ & Helical & \multicolumn{2}{c|}{} & $0^{\ast}$ & $2.32(3)^{\dagger}$ & $2.32(3)i^{\dagger}$ & $2.32(3)$ & $18.1$\tabularnewline
\hline 
 &  & $\mathbf{q}_{1}$ (r.l.u.) & $\mathbf{q}_{2}$ (r.l.u.) & $\mu_{1}$ ($\mu_{\mathrm{B}}$) & $\mu_{2}$ ($\mu_{\mathrm{B}}$) & $\mu_{3}$ ($\mu_{\mathrm{B}}$) &  & \tabularnewline
\hline 
$5$ & Multi-$\mathbf{q}$ & \multirow{1}{*}{$[0.2203(2),0,0]$} & \multirow{1}{*}{$[0,0.2251(4),0]$} & $5.45(3)$ & $4.63(5)$ & $0.65(17)$ & $7.18(5)$ & $17.6$\tabularnewline
\hline 
\end{tabular}\caption{\label{tab:table1}Fitted values of magnetic structure parameters
from Rietveld refinements. The derived maximum value of the ordered
magnetic moment per Gd is also given. Values denoted with a dagger
($\dagger$) were constrained to be of equal magnitude. Parameter
uncertainties indicate $1\sigma$ confidence intervals.}
\end{table*}

To obtain further insight into the temperature evolution of the magnetic
structure, we tested structure models against our POWGEN data using
magnetic Rietveld refinements. We consider first the data collected
in Phase 4 ($T=42$\,K) and structures of the form \citep{Wills_2001}
\begin{equation}
\text{\ensuremath{\mu}}(\mathbf{R})\propto\left(\mu_{\mathbf{a}},\mu_{\mathbf{b}},\mu_{\mathbf{c}}\right)\exp(-2\pi\mathrm{i}\mathbf{q}\cdot\mathbf{R})+\mathrm{c.c.},\label{eq:sine}
\end{equation}
where c.c. denotes the complex conjugate, $\mathbf{q}$ denotes the
propagation vector, $\mathbf{R}$ denotes a lattice vector, and $\mu_{\mathbf{a}},\mu_{\mathbf{b}},\mu_{\mathbf{c}}$
are (possibly complex) basis-vector components related to unit vectors
parallel to the crystallographic $\mathbf{a},\mathbf{b}$, and $\mathbf{c}$
axes. Initially, we consider an amplitude-modulated sine structure
with three real parameters $\mu_{\mathbf{a}},\mu_{\mathbf{b}},\mu_{\mathbf{c}}$,
corresponding to the maximum values of the magnetic moment along $\mathbf{a},\mathbf{b}$,
and $\mathbf{c}$, respectively. This refinement yields excellent
agreement with the $T=42$\,K experimental data with $R_{\mathrm{wp}}=18.0$\,\%;
fits to the full profile are shown in Figure~\ref{fig:fig2}(b),
fits to selected magnetic Bragg peaks in Figures~\ref{fig:fig2}(c)
and \ref{fig:fig2}(d), and a graphical representation of the refined
structure is shown in Figure \ref{fig:fig2}(e). Refined values of
the basis-vector components for all models are given in Table~\ref{tab:table1}.
The refined values of $\mu_{\mathbf{b}}$ and $\mu_{\mathbf{c}}$
are equal within error, while $\mu_{\mathbf{a}}$ is much smaller,
suggesting that a sine structure with magnetic moments polarized along
the $[011]$ direction is an appropriate single-parameter model. This
model yields a very similar fit quality ($R_{\mathrm{wp}}=18.1$\,\%)
to the three parameter model. Due to the effect of powder averaging,
however, the sine structure with $[011]$ moment direction is not
a unique solution: a helical magnetic structure with magnetic moments
confined to the $bc$ plane yields an identical fit. These models
are physically distinct: in particular, in the helical case, the magnetic
moment length is the same on every site in the crystal, whereas in
the sine case, the magnetic moment amplitude is modulated. The magnetic
moment does not exceed the maximum expected value of $7.0$\,$\mu_{\mathrm{B}}$
per Gd for either model, so both models are physically reasonable;
moreover, several examples are known where an amplitude-modulated
magnetic structure exists at elevated temperatures (e.g., \citep{Islam_1998,Blanco_2010,Mendive-Tapia_2019}).
Further experiments, such as single-crystal neutron diffraction, would
be needed to distinguish between the sine and helical possibilities.

At our base temperature of $5$\,K, we compare our powder neutron
diffraction data with the topological spin-stripe structure \citep{Wood_2023}
shown in Figure \ref{fig:fig2}(e). The moment orientations in this
model are a sum of three components \citep{Wood_2023},
\begin{align*}
\text{\ensuremath{\mu}}_{\mathrm{LT}}(\mathbf{R}) & \propto\left(0,\mu_{1},\mathrm{i}\mu_{1}\right)\exp(-2\pi\mathrm{i}\mathbf{q}_{1}\cdot\mathbf{R})\\
 & +\left(\mu_{2},0,0\right)\exp(-2\pi\mathrm{i}\mathbf{q}_{2}\cdot\mathbf{R})\\
 & +\left(0,\mu_{3},\mathrm{i}\mu_{3}\right)\exp(-2\pi\mathrm{i}\mathbf{q}_{3}\cdot\mathbf{R})+\mathrm{c.c.},
\end{align*}
which are a circular helix with moments in the $bc$ plane with propagation
vector $\mathbf{q}_{1}$, an amplitude-modulated sine structure with
moments along $[100]$ with propagation vector $\mathbf{q}_{2}$,
and a second circular helix with with propagation vector $\mathbf{q}_{3}=\mathbf{q}_{1}+2\mathbf{q}_{2}$.
The inclusion of $\mathbf{q}_{3}$ with $\mu_{3}=\mu_{2}^{2}/4\mu_{1}$
constrains the magnitude of the magnetic moment to be the same on
all sites in the crystal \citep{Wood_2023}. Our powder data provide
a particularly stringent test of this model, since their high resolution
allow peaks associated with $\mathbf{q}_{1}$ and $\mathbf{q}_{2}$
to be clearly resolved. First, we allow $\mu_{1}$, $\mu_{2}$, and
$\mu_{3}$ to refine independently, which yields excellent agreement
with the experimental data ($R_{\mathrm{wp}}=17.6\%$). The refined
values of $\mu_{1}$, $\mu_{2}$, and $\mu_{3}$ and $\mathbf{q}_{1}$
and $\mathbf{q}_{2}$ are in good agreement with Ref.~\citep{Wood_2023}
(see Table~\ref{tab:table1}). Peaks associated with $\mathbf{q}_{3}$
are not directly resolved, but the value of $\mu_{3}$ is constrained
by the overall goodness-of-fit. Notably, in our refinements, $\mathbf{q}_{1}$
and $\mathbf{q}_{2}$ are swapped compared to Ref.~\citep{Wood_2023},
so that the smaller wavevector corresponds to the helical modulation
and the larger wavevector to the sinusoidal one; this choice is necessary
to obtain satisfactory agreement with our data. Second, we constrain
the ratios $\mu_{2}/\mu_{1}$ and $\mu_{3}/\mu_{1}$ to be the same
as Ref.~\citep{Wood_2023}, and refine only the magnitude of $\mu_{1}$.
This procedure ensures that the magnetic moment is identical on all
sites in the crystal, and again yields good agreement with the data
($R_{\mathrm{wp}}=17.6\%$). The refined magnetic moment magnitude
of $6.98(2)$\,$\mu_{\mathrm{B}}$ per Gd is in excellent agreement
with the spin-only value of $7.0\,\mu_{\mathrm{B}}$ for spin-only
Gd$^{3+}$ions. Therefore, our refinements of high-resolution powder
neutron diffraction data provide strong support for the 2-\textbf{$\mathbf{q}$}
topological-spin-stripe ground state.

\begin{figure*}[t]
\includegraphics{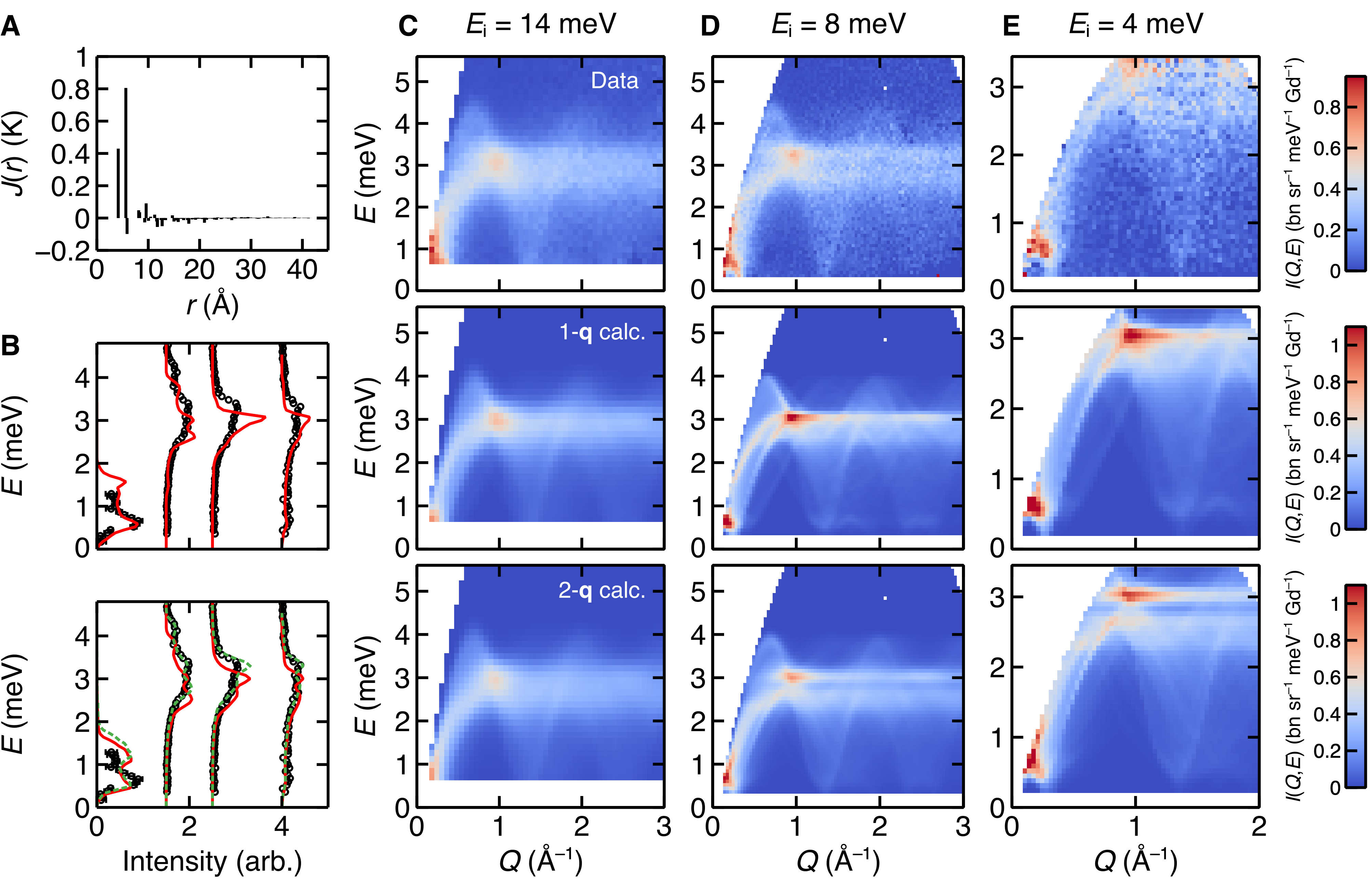}
\centering{}\caption{\label{fig:fig3}\textbf{Inelastic neutron scattering data and spin-wave
model calculations.} (a) Dependence of magnetic interactions $J(r)$
determined from DFT calculations \citep{Bouaziz_2022} on radial distance
$r$. Positive values indicate ferromagnetic interactions. (b) Constant-$Q$
cuts of inelastic neutron-scattering data compared with linear-spin-wave
calculations, showing cuts at (left to right) $Q=0.245$, $0.735$,
$0.980$, and $1.330\,\text{\ensuremath{\mathrm{\mathring{A}}^{-1}}}$.
Cuts are from $E_{i}=4$\,meV ($Q=0.245\,\text{\ensuremath{\mathring{A}^{-1}}}$)
and $8$\,meV data (all other cuts) and the $Q$-resolution is $\pm0.0175\,\text{\ensuremath{\mathrm{\mathring{A}}^{-1}}}$.
Calculations are for $J(r)$ and the 1-$\mathbf{q}$ helical structure
(red lines in upper panel), $J(r)$ and the 2-$\mathbf{q}$ topological
spin stripe structure (red lines in lower panel), and $1.1J(r)$ in
the 2-$\mathbf{q}$ topological spin stripe structure (dotted green
lines in lower panel). (c) Data with $E_{i}=14$\,meV (top panel),
1-$\mathbf{q}$ spin-wave calculation (middle panel), and 2-$\mathbf{q}$
calculation (lower panel). (d) Data with $E_{i}=8$\,meV (top panel),
1-$\mathbf{q}$ spin-wave calculation (middle panel), and 2-$\mathbf{q}$
calculation (lower panel). (e) Data with $E_{i}=4$\,meV (top panel),
1-$\mathbf{q}$ spin-wave calculation (middle panel), and 2-$\mathbf{q}$
calculation (lower panel). All calculations are convolved with the
instrumental energy resolution of $0.52$, $0.23$, and $0.13$\,meV
for $E_{i}=14$, $8$, and $4$\,meV, respectively, from Gaussian
fits to the elastic line of the experimental data. }
\end{figure*}

With a model of the magnetic ground-state structure in hand, we now
investigate the low-energy spin dynamics. Traditionally, such analysis
proceeds by optimizing a few magnetic exchange parameters against
inelastic neutron scattering data. However, this approach provides
at best an effective description of intermetallic magnets such as
GdRu$_{2}$Si$_{2}$, because the long-ranged nature of the RKKY interactions
implies that an impractically large number of interaction parameters
must be optimized for a complete description. For this reason, we
instead compare our inelastic neutron-scattering data with predictions
of the density-functional-theory (DFT) simulations reported in Ref.~\citep{Bouaziz_2022},
which provide a qualitatively correct description of the magnetic
field\emph{ vs. }temperature phase diagram of GdRu$_{2}$Si$_{2}$
\citep{Bouaziz_2022}. The DFT interactions describe an isotropic
(Heisenberg) Hamiltonian,
\begin{align*}
H_{\mathrm{iso}} & =-\sum_{i>j}J_{ij}\mathbf{S}_{i}\cdot\mathbf{S}_{j},
\end{align*}
where $\mathbf{S}_{i}$ denotes a spin vector with position $\mathbf{R}_{i}$
and quantum number $S=7/2$, and $J_{ij}\equiv J(\mathbf{R}_{j}-\mathbf{R}_{i})$
is the interaction between spins at $\mathbf{R}_{i}$ and $\mathbf{R}_{j}$.
The dependence of the DFT interactions on radial distance, $r=|\mathbf{R}_{j}-\mathbf{R}_{i}|$,
is shown in Figure~\ref{fig:fig3}(a). While the magnitudes of the
interactions decay rapidly with increasing distance, treating their
long-ranged nature is nevertheless crucial: A maximum distance of
$10a=41.6$\,\AA is needed to obtain a calculated magnetic propagation
vector $\mathbf{q}_{\mathrm{calc}}=[0.184,0,0]$ in reasonable agreement
with the experimental value, whereas truncating the interactions at
a distance of $3a$ yields an incorrect ferromagnetic propagation
vector $\mathbf{q}_{\mathrm{calc}}=\mathbf{0}$. The long-range nature
of the interactions hints towards an RKKY mechanism \citep{Bouaziz_2022}.

Figure~\ref{fig:fig3}(b--e) compares our inelastic neutron-scattering
data with calculations of the magnetic excitation spectra performed
using linear spin-wave theory. Figure~\ref{fig:fig3}(b) shows constant-$Q$
cuts of the experimental data, while the top panels of Figure~\ref{fig:fig3}(c),
(d), and (e) show data collected with $E_{i}=4$, $8$, and $14$\,meV,
respectively. Our data show an overall excitation bandwidth of approximately
$4.5$\,meV. The $E_{i}=4$ meV data clearly show a region of high
intensity at $0.5\lesssim E\lesssim1.0$\,meV at small $Q$, while
the $E_{i}=8$ meV data suggest that the spectral weight at $E\sim3$\,meV
consists of two nearly-dispersionless excitation bands at $E\approx2.8$
and $3.2$\,meV. To model our inelastic neutron-scattering data,
we calculate the neutron-scattering spectrum using linear spin-wave
theory, as implemented in the SpinW program \citep{Toth_2015}. The
Hamiltonian is the sum of the isotropic interactions calculated by
DFT (see Figure~\ref{fig:fig3}(a)) and the long-ranged dipolar interaction,
so that the total Hamiltonian is given by
\[
H=H_{\mathrm{iso}}+D\sum_{i>j}\frac{\mathbf{S}_{i}\cdot\mathbf{S}_{j}-3\left(\mathbf{S}_{i}\cdot\hat{\mathbf{r}}_{ij}\right)\left(\mathbf{S}_{j}\cdot\hat{\mathbf{r}}_{ij}\right)}{\left(r_{ij}/r_{1}\right)^{3}},
\]
where $D=0.047$\,meV is the energy scale of the dipolar interaction
at the nearest-neighbor distance $r_{1}$, which is proportional to
the squared magnetic moment and is thus relatively large for Gd$^{3+}$.
We also include a small easy-axis single-ion anisotropy term, $\Delta=-0.005$
meV, to stabilize magnetic moments in the $bc$ plane. 

Importantly, linear spin-wave theory requires a magnetically ordered
state that is a local energy minimum of the assumed magnetic Hamiltonian.
If there is more than one such state, each may have a distinct magnetic
excitation spectrum \citep{Paddison_2021}. In GdRu$_{2}$Si$_{2}$,
the magnetic Gd ions occupy a Bravais lattice, and the DFT calculation
includes Heisenberg magnetic interactions only. In this case, the
magnetic ground state of the Hamiltonian is a 1-$\mathbf{q}$ helix.
This result is not fully consistent with the experimental result from
neutron diffraction, which shows that the observed magnetic ground
state is actually a 2-$\mathbf{q}$ state. However, since the largest
basis-vector component in the 2-$\mathbf{q}$ structure describes
a helix, and the neutron-scattering intensity is proportional to the
square of the basis-vector component magnitude, we expect that scattering
from the helical component will make the largest contribution to the
observed intensity. The calculated spin-wave spectrum of the helical
state is shown in the upper panel of Figure~\ref{fig:fig3}(b) and
the middle panels of Figure \ref{fig:fig3}(c), (d), and (e). The
overall energy scale agrees well with the experimental data and the
main features of the data are well reproduced. This level of agreement
is remarkable, considering that the comparison relies on independent
first-principles calculations \citep{Bouaziz_2022} and does not include
any free parameters. 

We now consider the excitation spectrum of the 2-$\mathbf{q}$ topological-spin-stripe
state \citep{Wood_2023}. Compared to the 1-$\mathbf{q}$ helical
state, this calculation introduces two complexities. First, to account
for the multi-$\mathbf{q}$ state, it is necessary to use a supercell
of the crystallographic unit cell; we consider a $5\times5\times1$
supercell containing 50 Gd$^{3+}$ ions, which enforces propagation
vectors of $[\frac{1}{5},0,0]$ and $[0,\frac{1}{5},0]$. Second,
the 2-$\mathbf{q}$ ground state has an energy $\approx2.8\%$ greater
than that of the helical state; hence, it is not an exact local energy
minimum of the Hamiltonian. These two limitations imply that the spin-wave
energies contain some imaginary values, which are unphysical and must
be discarded in the calculation of the spin-wave intensities \citep{Bai_2019}.
Since only a small fraction ($\approx0.06\%$) of the spin-wave modes
considered have imaginary eigenvalues, we expect that the calculated
spectrum still represents a reasonable approximation to the spin-wave
spectrum of the 2-$\mathbf{q}$ ground state. The excitation spectrum
of the 2-$\mathbf{q}$ topological-spin-stripe state is shown in the
upper panel of Figure~\ref{fig:fig3}(b), and the lower panels of
Figure~\ref{fig:fig3}(c), (d), and (e). It shows clearly improved
agreement with the experimental data compared to the 1-$\mathbf{q}$
helical calculation. Notably, the 2-$\mathbf{q}$ calculation reproduces
the two nearly-dispersionless excitation bands at $E\approx2.8$ and
$3.2$\,meV that are not reproduced by the helical calculation. The
agreement with experiment is further improved by scaling the energies
of all the interactions by a factor of $1.1$. This scaling provides
essentially quantitative agreement between the 2-$\mathbf{q}$ calculation
and the experimental data, as shown by the constant-$Q$ cuts in the
lower panel of Figure~\ref{fig:fig3}(b). It also yields improved
agreement between the experimental Weiss temperature ($41$\,K) and
the calculated value obtained as $\theta_{\mathrm{calc}}=\frac{1}{3}S(S+1)\sum_{j}Z_{j}J_{0j}$,
which is $28$\,K after rescaling.

Our neutron-scattering experiments provide an essential quantitative
test of the magnetic interactions of GdRu$_{2}$Si$_{2}$ calculated
from first-principles theory. Our results show that the theoretical
calculations \citep{Bouaziz_2022} provide an accurate description
of our experimental data, providing confirmation of the methodology
and the nature of the interactions. In particular, our study indicates
that the magnetic interactions of GdRu$_{2}$Si$_{2}$ are dominated
by isotropic (Heisenberg) exchange, and the long-ranged nature of
the DFT interactions strongly supports interpretations in terms of
a generalized RKKY mechanism \citep{Bouaziz_2022,Paddison_2022,Nomoto_2023}.
This mechanism is supported by quantum-oscillation measurements in
GdRu$_{2}$Si$_{2}$ \citep{Matsuyama_2023}, but is not universal
in magnetic intermetallics, where superexchange can dominate \citep{Bouaziz_2024}.
The excellent agreement of first-principles calculations with experimental
data also shines light on the few areas in which the theory does \emph{not
}fully agree with experiment. Most importantly, our experimental results
confirm the 2-$\mathbf{q}$ nature of the magnetic ground state, but
this state cannot be stabilized by purely isotropic interactions;
an anisotropic or multi-spin contribution is necessary to stabilize
it; two possible candidates are a biqudratic interaction \citep{Ozawa_2016}
or a bond-dependent anisotropy \citep{Hayami_2021}. However, the
excellent agreement of the purely isotropic model with our spectroscopic
model suggests that the anisotropic contribution is much smaller than
the isotropic one. As such, the determination of this contribution
represents an open challenge for theory and experiment alike, which
motivates further experiments on single-crystal samples \citep{Hayami_2021}.

\acknowledgements{We are grateful to Geetha Balakrishnan (Warwick), Eleanor Clements
(ORNL), Shang Gao (ORNL), and George Wood (Warwick) for valuable discussions.
This work was supported by the U.S. Department of Energy, Office of
Science, Basic Energy Sciences, Materials Sciences and Engineering
Division, and used resources at the High Flux Isotope Reactor and
Spallation Neutron Source, DOE Office of Science User Facilities operated
by the Oak Ridge National Laboratory. The isotope used in this research
was supplied by the U.S. Department of Energy Isotope Program, managed
by the Office of Isotope R\&D and Production.}

\end{document}